# Ilmenite-Type Ca$_x$IrO$_3$ via Topochemical Ion Exchange: Stacking Faults and Low-Temperature Magnetic Anomaly

Haruki Kira[1], Yuya Haraguchi[1,*], Wataru Yokoshima[1], Daisuke Nishio-Hamane[2], and Hiroko Aruga Katori[1]

[1]*Department of Applied Physics and Chemical Engineering, Tokyo University of Agriculture and Technology, Koganei, Tokyo 184-8588, Japan*

[2]*The Institute for Solid State Physics, The University of Tokyo, Kashiwa, Chiba 277-8581, Japan*

We report the synthesis of an ilmenite-type polymorph of Ca$_x$IrO$_3$ distinct from the known post-perovskite and perovskite phases, via low-temperature topochemical Ca$^{2+}$/2Na$^+$ exchange from Na$_2$IrO$_3$. Powder X-ray diffraction is indexable in $R\bar{3}$, and whole-pattern modelling that includes layer-glide faults indicates that the selective broadening can be captured by a first-order Markov stacking description based on stochastic switching between two symmetry-equivalent lateral stacking steps, with explicit model dependence and an uncertainty of at least several percent. A freezing-like bulk magnetic anomaly is suggested at $T^* \approx 25$ K (defined by the onset of a ZFC/FC bifurcation at $\mu_0 H = 10$ mT), accompanied by a broad heat-capacity feature and Curie–Weiss behavior with a large negative Weiss temperature of $\theta_\mathrm{W} \approx -98$ K. The effective moment $\mu_\mathrm{eff} = 1.68\ \mu_\mathrm{B}$ per Ir is consistent with $J_\mathrm{eff} = 1/2$ for an Ir$^{4+}$. SEM-EDX suggests an A-site content below unity (Ca/Ir < 1); accordingly, we describe the ion-exchanged product using the nonstoichiometric formula Ca$_x$IrO$_3$. These results identify ilmenite-type CaIrO$_3$ as a honeycomb iridate in which stacking disorder can be quantified (with caveats regarding model and instrument correlations) and related to its low-temperature magnetic behavior.

## I. INTRODUCTION

Iridium oxides based on 5d electrons have drawn sustained interest because the competition between strong spin–orbit coupling (SOC) and electron correlations yields a low-energy manifold in which the $J_\mathrm{eff} = 1/2$ state acts as the relevant degree of freedom [1,2]. In this setting, bond-dependent interactions of the Kitaev type can become prominent [3,4]. Such SOC-entangled electronic states are highly sensitive to crystallographic degrees of freedom: not only layering and local coordination distortions but also stacking faults arising from in-plane layer translations can govern both the presence of long-range order and the characteristic energy scales.

Even within the simple stoichiometry $A$IrO$_3$, the diversity of crystal structures is striking. For larger $A$-site cations, both three-dimensional perovskite and hexagonal perovskite polymorphs tend to be stabilized. A representative case is SrIrO$_3$, widely known as a distorted perovskite in which strong SOC acting on a three-dimensional Ir–O–Ir network produces a semimetal that nonetheless retains signatures of electronic correlations [5,6]. In the nearly undistorted perovskite obtained under high pressure, Dirac-like band crossings and non-Fermi-liquid transport have been discussed [7]. By contrast, BaIrO$_3$ commonly adopts a hexagonal polymorph containing face-sharing units; there the interplay of SOC and trimerization leads to a charge-order-like transition and a weak ferromagnetic response, highlighting the coupled roles of correlations and lattice degrees of freedom [8-10]. Recent advances in high-pressure synthesis and thin-film stabilization have broadened access to perovskite polymorphs with reduced octahedral distortion, for which indicators of three-dimensional electronic states, in sharp contrast to the hexagonal polytypes, are being reported [11-13]. The comparison between SrIrO$_3$ and BaIrO$_3$ thus illustrates how crystallographic freedom under strong SOC sets the electronic properties, and how high-pressure synthesis can provide low-distortion reference phases that sharpen the discussion of underlying physics.

With an intermediate $A$-site radius, CaIrO$_3$ crystallizes at ambient pressure in the post-perovskite structure. Its antiferromagnetic order has been described within a framework that includes anisotropic exchange among $J_\mathrm{eff} = 1/2$ pseudospins, incorporating Kitaev-type bond dependence [14,15]. In contrast, the perovskite polymorph of CaIrO$_3$ stabilized under high pressure has been identified as a semimetal with Dirac-like band crossings arising from the combination of strong SOC and crystal symmetry, making clear that polymorphism alone can drastically change the electronic and magnetic ground states even at fixed composition [16,17]. This sensitivity of the structure–property relationship motivates a comparison with phases that more readily appear for smaller $A$, such as ilmenite, to assess how crystallographic freedom and defects influence Kitaev-related behavior.

As ionic radii of $A$ decrease further, corundum-derived ilmenite structures are generally favored [18,19]. In iridium oxides, however, the ilmenite polymorph is only weakly stable at ambient pressure and conventional high-temperature reactions tend to yield more stable competing phases. Consequently, synthetic strategies that selectively capture metastable states while preserving the oxygen framework are required Low-temperature topochemical reactions that exchange only the cations [20,21], as well as molten-salt routes, have proven effective in this regard. Because these approaches allow gentle control of lattice strain and layer order, they are also well suited for testing how lattice degrees of freedom correlate with magnetism.

The ilmenite structure hosts transition-metal ions on a two-dimensional honeycomb network and serves as a common platform for honeycomb-lattice magnets such as $MgMnO_3$ [22] and $CoTiO_3$ [23]. In ilmenite-type iridates, strong SOC activates bond-dependent interactions, bringing these systems close to the Kitaev model [24-26]. Ilmenites also tend to accommodate stacking faults due to layer glides [27]. Small shifts or rotations of the coordination polyhedral can substantially modify the SOC-entangled Ir states. As a result, even at fixed stoichiometry one may encounter a range of magnetic and thermal responses tied to structural variability, and controlling defect density and layer order becomes a practical handle for property design.

Across several Kitaev-related compounds, stacking faults have already been implicated as a key factor setting up the ground state. In the honeycomb system $Ag_3LiIr_2O_6$, samples with higher fault density show suppressed long-range order, whereas reducing faults leads to the emergence of order [21,28]. The proton-exchanged $H_3LiIr_2O_6$ accommodates large interlayer shifts and behaves as a spin-liquid candidate, though the extent to which defect-derived contributions enter its continuum excitations and thermal response remains under discussion [29-31]. For ilmenite-type $MgIrO_3$, $ZnIrO_3$, and $CdIrO_3$, magnetic order is observed despite conspicuous stacking faults, but the specific-heat anomalies are markedly broadened, pointing to a partial suppression of three-dimensional correlations [24,25]. These observations call for a quantitative assessment, spanning different structure families, of how stacking faults move the boundary between ordered and disordered states.

To make progress, it is important to go beyond single-compound case studies and track systematic trends with a parameter that cuts across materials, such as the $A$-site ionic radius. Choosing $A$-ion larger than Cd enables tests of how Ir–O–Ir bond angles, interlayer correlations, and defect tolerance co-vary. $CaIrO_3$ is a natural candidate. However, at ambient pressure the post-perovskite structure is frequently reported and is thermodynamically favored [14,15]. Therefore, targeting the ilmenite polymorph of $CaIrO_3$ requires a low-temperature topochemical cation-exchange route that preserves the anion framework and selectively traps the metastable phase. More generally, topochemical transformations, including ion exchange and low-temperature redox reactions, are widely used to access metastable oxide frameworks by retaining the parent anion sublattice and crystallographic registry while avoiding high-temperature equilibration pathways [32-36]. Aliovalent divalent-for-monovalent ion-exchange routes also have early precedents, for example the ilmenite variety of $CaSnO_3$ reported by Durand and Loiseleur [37].

Here we synthesize ilmenite-type $Ca_xIrO_3$ at ambient pressure via a low-temperature topochemical reaction using $Na_2IrO_3$ as a precursor, using an aliovalent $Ca^{2+}/2Na^+$ exchange conceptually analogous to divalent-for-monovalent topotactic ion-exchange reactions [38,39], and we show that despite pronounced stacking faults generated by layer glides, a freezing-like bulk magnetic anomaly appears at approximately 25 K, as established by a ZFC/FC bifurcation at low field together with a broad heat-capacity feature. Notably, simple ionic-size considerations place stoichiometric $CaIrO_3$ among the least favorable candidates for an ilmenite-type honeycomb framework in the $AIrO_3$ series; therefore, the ability to reach an ilmenite-type polymorph at low temperature, albeit with strong stacking disorder, is a defining crystal-chemistry aspect of the present work. We therefore refrain from claiming unambiguous long-range order and instead focus on correlating stacking disorder with the low-temperature anomaly. These results identify ilmenite-type iridium oxides, including $Ca_xIrO_3$, as practical systems for probing the correlation between stacking faults and magnetism in the context of the Kitaev model.

## II. EXPERIMENTAL METHODS

**Precursor synthesis.** $Na_2IrO_3$ was prepared by a conventional solid-state route [40]. Stoichiometric amounts of dried $Na_2CO_3$ and Ir were homogenized (with a 10 wt% excess of $Na_2CO_3$ to compensate sodium volatility), pressed into pellets, and fired in air at 700 °C for 12 h followed by 800 °C for 48 h. The pellets were furnace cooled to 400 °C and then quenched.

**Table I.** Crystallographic parameters for $Ca_xIrO_3$ (space group $R\bar{3}$) determined by Rietveld refinement of powder SXRD data at 300 K. Average structural model used as input for the stacking-fault analysis. The refined lattice parameters are $a = 5.354618(12)$ Å and $c = 15.23152(5)$ Å. $B$ denotes the atomic displacement parameter.

|    | Site | $x$       | $y$       | $z$          | $B$ (Å)  |
|----|------|-----------|-----------|--------------|----------|
| Ca | 6c   | 0         | 0         | 0.1356(1)    | 0.30(4)  |
| Ir | 6c   | 0         | 0         | 0.337541(17) | 0.240(3) |
| O  | 18f  | 0.3835(4) | 0.0131(4) | 0.07054(9)   | 0.48(3)  |

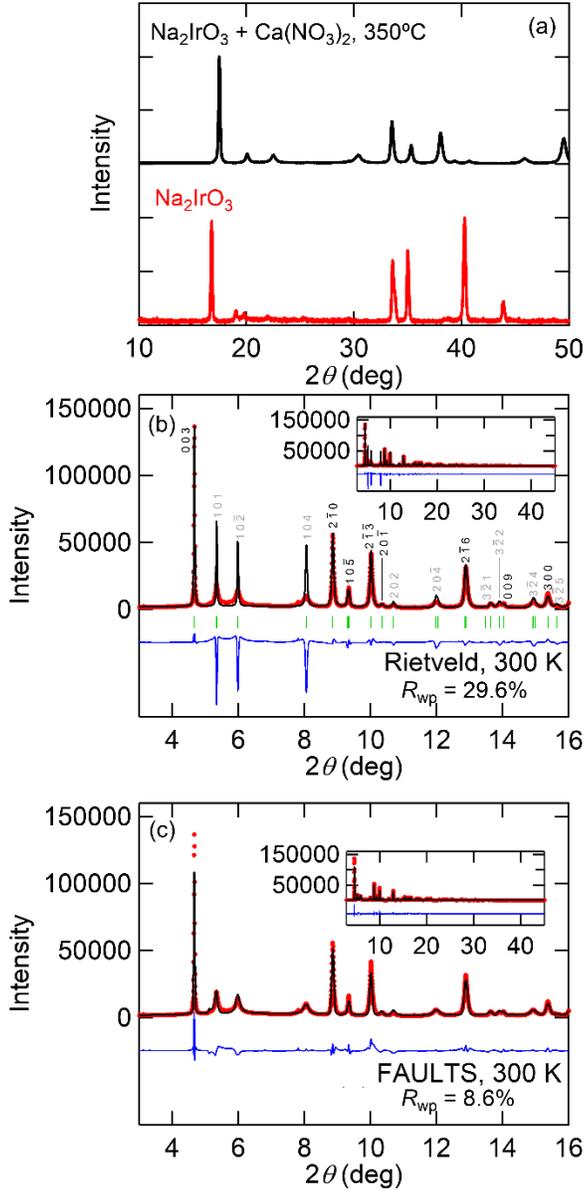

**Fig. 1** (a) Laboratory powder X-ray diffraction (XRD) patterns collected with Cu Kα radiation for the Na$_2$IrO$_3$ precursor (red) and the product obtained after reacting Na$_2$IrO$_3$ with Ca(NO$_3$)$_2$ at 350 °C (black). The patterns are vertically offset for clarity. (b, c) Synchrotron powder XRD pattern of the CaIrO$_3$ product measured at 300 K at SPring-8 BL13XU (lambda = 0.413463 angstrom), shown with (b) a conventional Rietveld refinement in space group $R\bar{3}$ and (c) a FAULTS refinement including stacking disorder. Observed, calculated, and difference profiles are shown in red, black, and blue, respectively. Tick marks indicate the calculated reflection positions for the average structure. Insets show the same datasets over a wider 2θ range.

**Table II.** Atomic coordinates of the basic stacking layer used in the FAULTS refinement for CaIrO3. The layer model corresponds to the single honeycomb slab shown in Fig. 2(a). It is defined in the hexagonal setting with cell parameters $a = b = 5.354618$ Å, $c = 5.077173$ Å, $\gamma = 120°$, and Laue symmetry -1. The coordinates define the stacking layer unit used in FAULTS; they do not represent a unit cell.

| Atom | x | y | z | B(Å) |
|---|---|---|---|---|
| Ir1 | 0 | 0 | 0.48216 | 0.24 |
| Ir2 | 2/3 | 1/3 | 0.51783 | 0.24 |
| Ca1 | 1/3 | 2/3 | 0.09309 | 0.30 |
| Ca2 | 0 | 0 | 0.90693 | 0.3 |
| O1 | 0.94983 | 0.29626 | 0.29439 | 0.48 |
| O2 | 0.70374 | 0.65357 | 0.29439 | 0.48 |
| O3 | 0.34643 | 0.05017 | 0.29439 | 0.48 |
| O4 | 0.71684 | 0.03707 | 0.7056 | 0.48 |
| O5 | 0.96293 | 0.67976 | 0.7056 | 0.48 |
| O6 | 0.32024 | 0.28316 | 0.7056 | 0.48 |

To suppress Na loss arising from reactions with atmospheric moisture, the product was immediately ground, re-pelletized with the same Na$_2$CO$_3$ excess, and a second anneal was carried out in air at 825 °C for 48 h, followed by furnace cooling to 400 °C. The resulting Na$_2$IrO$_3$ powders were stored in an Ar-filled glovebox.

**Topochemical synthesis.** The ilmenite type Ca$_x$IrO$_3$ was targeted via the following ideal cation exchange reaction,

$$Na_2IrO_3 + Ca(NO_3)_2 \rightarrow CaIrO_3 + 2NaNO_3 \quad (1)$$

Stoichiometric amounts of Na$_2$IrO$_3$ and Ca(NO$_3$)$_2$ were weighed, thoroughly mixed, pressed into pellets, sealed in a Pyrex tube under Ar, and heated at 350 °C for 48 h. Under these conditions, NaNO$_3$ (m.p. 308 °C) forms as a molten phase that promotes ion mobility and enables topochemical Ca$^{2+}$/Na$^+$ exchange while preserving the IrO$_6$ framework. The reacted solid was washed with deionized water to remove NaNO$_3$, and dried to yield Ca$_x$IrO$_3$ powder.

During method development, we also examined larger Ca(NO$_3$)$_2$ loadings. However, using a large excess of Ca(NO$_3$)$_2$ led to additional unidentified reflections after washing, whereas the lattice parameters of the CaIrO$_3$ main phase did not show a systematic dependence on the Ca(NO$_3$)$_2$ amount within our resolution. We therefore focus on the

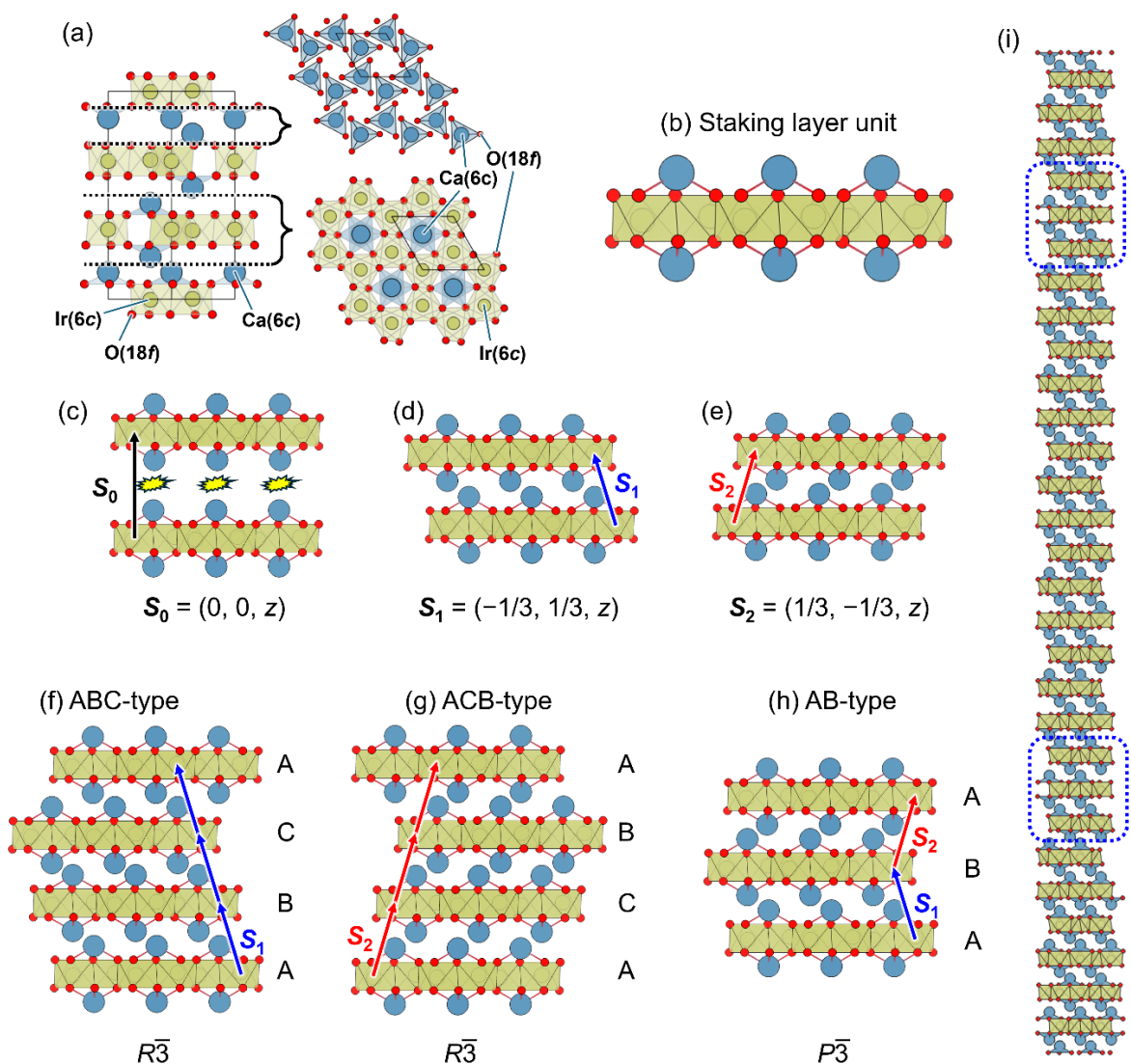

**Fig. 2** (a) Ilmenite-type crystal structure of CaIrO$_3$, highlighting the description in terms of quasi-rigid honeycomb layers stacked along the hexagonal $c$ direction (side view and in-plane projection). Ca and Ir occupy the 6$c$ sites and O occupies the 18$f$ site (as labeled); IrO$_6$ octahedra are shown in polyhedral representation. (b) Definition of the stacking-layer unit (a single quasi-rigid honeycomb slab) used for the stacking description. (c–e) Candidate interlayer translation vectors relating adjacent layers: the eclipsed stacking $S_0 = (0, 0, z)$, which produces unacceptably short Ca–Ca contacts (schematically indicated), and two symmetry-related in-plane glides $S_1 = (-1/3, 1/3, z)$ and $S_2 = (1/3, -1/3, z)$ that preserve layer connectivity. (f–h) Representative long-range stacking sequences generated by repeating $S_1$ (ABC-type) or $S_2$ (ACB-type), or by alternating $S_1$ and $S_2$ (AB-type); the corresponding average symmetries are indicated. (i) Example of a faulted stacking sequence in which stacking faults are randomly incorporated via stochastic switches between $S_1$ and $S_2$; dashed boxes highlight locally faulted segments.

stoichiometric condition in this work to minimize secondary phases and to provide a consistent basis for the structural analysis.

**Structural characterization.** The structures of the precursor and final products were initially checked by an X-ray powder diffractometer (MiniFlex600, Rigaku, Cu K$\alpha$), and chemical analysis was conducted using a scanning electron microscope (JEOL IT100) equipped with an energy dispersive x-ray spectroscope (EDX) with 15 kV, 0.8 nA, 1 μm beam diameter). Because EDX requires a flat surface and a locally homogeneous interaction volume (typically on the order of a few micrometers) for reliable quantification, the exchanged powder was first compacted into a dense pellet using a diamond anvil press and the pellet surface was polished to create flat regions. EDX spectra were then collected from multiple polished areas to minimize topography-related artifacts. For structural refinement, in-situ synchrotron XRD (SXRD) data were collected at BL13XU beamline of SPring-8 using a wavelength of λ = 0.413606 Å, with the samples sealed in a 0.3 mm-diameter quartz capillary. The data collected under these conditions were used for Rietveld structural refinements of the involved crystalline phases by means of the Z-RIETVELD software [42]. Then the resulting parameters were used as input for Rietveld refinement including the effect of stacking faults using the FAULTS software [43].

**Physical property measurements.** Magnetization was measured using an MPMS (sample mass: 24.75 mg) under ZFC/FC protocols from 1.8 to 300 K under magnetic fields up to 7 T, and specific heat was measured using a PPMS (sample mass: 14.016 mg) by the relaxation method from 2 to 300 K under fields up to 5 T. For the relaxation measurements, the powder sample was lightly cold pressed into a pellet to improve thermal contact with the measurement platform. All measurements were carried out at the Institute for Solid State Physics, The University of Tokyo.

### III. RESULTS

#### A. Crystal Structure

Powder SXRD was used to examine the crystal structure of $Ca_xIrO_3$. The laboratory Cu Kα XRD patterns in Fig. 1(a) compare the $Na_2IrO_3$ precursor with the sample after the topochemical reaction with $Ca(NO_3)_2$ at 350°C, and the clear change in diffraction peak positions (with the $Na_2IrO_3$ peaks being replaced by a new set of reflections) evidences that the ion-exchange reaction proceeds and yields the $Ca_xIrO_3$ phase. Figure 1(b) shows the measured pattern together with a conventional Rietveld refinement in space group $R\bar{3}$ (Table I). All reflections are indexable within the ilmenite-type metric, but the peak shapes are highly non-uniform: a subset of reflections remains sharp within the instrumental resolution, whereas many general *hkl* reflections exhibit pronounced, anisotropic broadening. In particular, *hk0* reflections and the allowed *00l* reflections remain comparatively sharp, while reflections that probe the lateral registry between adjacent layers are selectively broadened. This signature is characteristic of stacking disorder produced by in-plane layer glides. Accordingly, a single ideal $R\bar{3}$ model captures the average structure but cannot reproduce the full line-shape distribution of the powder profile.

To interpret this behavior in a chemically transparent way, we adopt a layer-stacking description analogous to that used for stacking-fault analysis in ilmenite-type $SnTiO_3$ [44]. The ilmenite framework can be viewed as a stack of quasi-rigid honeycomb layers formed by edge-sharing $IrO_6$ octahedra (Fig. 2(a)). In this layered view, Ca ions occupy the sites above and below the honeycomb voids. Importantly, the Ca sublattice is strongly buckled: each Ca ion is displaced toward one side of the neighboring IrO6 sheet, giving a highly asymmetric local environment that can be pictured as a short-bond, pseudo-trigonal-pyramidal motif embedded within an overall sixfold coordination. These oppositely oriented Ca pyramids are nested across adjacent layers (Fig. 2(a)), highlighting that the essential structural unit for discussing disorder is a quasi-rigid honeycomb layer together with its associated Ca environment. The corresponding stacking-layer unit used in the following analysis is shown in Fig. 2(b).

In hexagonal coordinates, the relative placement of one layer on the next can be described by a translation vector consisting of an in-plane shift and an interlayer step along *c*. As illustrated in Fig. 2(c), the eclipsed stacking $S_0 = (0, 0, z)$ is not permitted because it would place Ca ions at unacceptably short separations; therefore, the lateral shifts $S_1 = (−1/3, 1/3, z)$ and $S_2 = (1/3, −1/3, z)$, which maximize Ca–Ca separation, are selected as the relevant stacking vectors. For the honeycomb layer geometry, there are two symmetry-related lateral translations that preserve the layer connectivity: $S_1 = (−1/3, 1/3, z)$ and $S_2 = (1/3, −1/3, z)$, where z denotes the layer-to-layer step (Figs. 2(d) and 2(e)). Repeating $S_1$ generates an ABC-type stacking sequence, while repeating $S_2$ generates an ACB-type stacking sequence (Figs. 2(f) and 2(g)). Importantly, these two long-range stackings are symmetry-equivalent in $R\bar{3}$. To model stacking disorder, we allow stochastic switches between $S_1$ and $S_2$ along the stacking direction, which randomly embed local fault motifs (including AB-type fragments) within the otherwise rhombohedral stacking, as illustrated in Fig. 2(i).

Within this framework, stacking disorder can be introduced in a minimal way as local switches between $S_1$-type and $S_2$-type steps. When the stacking vector switches ($S_1$ to $S_2$, or $S_2$ to $S_1$), the local sequence necessarily passes through an AB-type registry (Fig. 2(h)). This AB-type motif is the simplest deviation from a purely rhombohedral sequence and can be generated by layer glides without invoking any rotation of the honeycomb layer building block. In other words, rather than asserting a unique long-range 2H polytype, we describe the disorder as predominantly rhombohedral stacking with a finite density of $S_1/S_2$ switching events that introduce short AB-type segments.

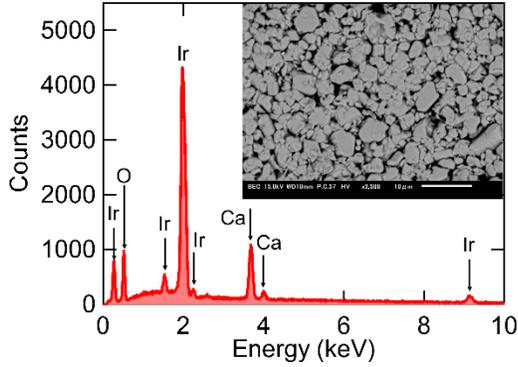

**Fig. 3** SEM-EDX characterization of the ion-exchanged $Ca_xIrO_3$ powder. Representative EDX spectrum (red) collected at an accelerating voltage of 15 kV from the product particles, showing characteristic peaks of O, Ca, and Ir. No distinct peak is observed at the energy of the Na K-alpha line (~1.04 keV) above background in this representative spectrum. Inset: backscattered-electron SEM image of the powder (scale bar: 10 μm).

**Table III**. Crystal structure parameters of $A$IrO$_3$ ($A$ = Mg, Zn, Cd, Ca, Sr, Ba). $N$ denotes the coordination number of the A site; $r_A$ is the ionic radius of the $A$ site ion; $t_p$ is the Goldschmidt tolerance factor for the perovskite structure; $t_i$ is the tolerance factor for the ilmenite structure, $\Delta e$ denotes the electronegativity difference between the A ion and Ir [44]. The abbreviations for the structure type (str.) are I = ilmenite, P = perovskite, PP = post-perovskite, and HP = hexagonal perovskite.

| Cation | $N$ | $r_A$(Å) | $t_p$ | $t_i$ | $\Delta e$ | str. | reference |
|---|---|---|---|---|---|---|---|
| $Mg^{2+}$ | 6 | 0.72 | 0.74 | 1.12 | 0.89 | I | [24,26] |
| $Zn^{2+}$ | 6 | 0.74 | 0.75 | 1.11 | 0.55 | I | [24,26] |
| $Cd^{2+}$ | 6 | 0.95 | 0.82 | 0.99 | 0.51 | I | [25] |
| $Ca^{2+}$ | 6 | 1.00 | 0.84 | 0.96 | 1.20 | I | This work |
| $Ca^{2+}$ | 8 | 1.12 | 0.88 | 0.91 | 1.20 | PP | [14] |
| $Ca^{2+}$ | 12 | 1.34 | 0.96 | 0.82 | 1.20 | P | [17] |
| $Sr^{2+}$ | 12 | 1.44 | 0.99 | 0.79 | 1.25 | P | [6,7] |
| $Ba^{2+}$ | 12 | 1.61 | 1.05 | 0.74 | 1.31 | HP | [12] |

For a quantitative description, we refined the powder profile using the program FAULTS by representing the stacking as a one-dimensional first-order Markov process for the sequence of $S_1$ and $S_2$ steps [43]. The stacking layer used for this refinement is defined by the atomic coordinates listed in Table II and illustrated in Fig. 2(a). These coordinates describe a single honeycomb layer unit within the hexagonal setting. The refinement yields transition probabilities that favor repeating the same stacking vector, $P(S_i$ to $S_i) \sim 0.87$, with a significant minority of switching events, $P(S_i$ to $S_j) \sim 0.13$ ($i \neq j$), for both $i = 1$ and 2. The corresponding calculated profile (Fig. 1(c)) reproduces the selective broadening and intensity redistribution far better than the ideal $R\bar{3}$ refinement. Because powder diffraction primarily constrains low-order stacking correlations and the extracted probabilities correlate with profile parameters such as instrumental resolution and anisotropic size/strain broadening, these numerical values should be regarded as approximate and model-dependent rather than a unique microscopic reconstruction of the stacking sequence. Nonetheless, the robust conclusion is that frequent layer-glide faults are required to account for the observed index-dependent broadening in $CaIrO_3$.

While incomplete exchange could in principle contribute to disorder, the selective broadening is consistently reproduced by a stacking-fault model based on switching between two symmetry-equivalent lateral stacking steps; the present diffraction analysis therefore focuses on stacking disorder rather than assigning a unique microscopic origin. Minor residual misfit remains, which we attribute to the limited peak-shape flexibility and the lack of preferred-orientation corrections within FAULTS; our analysis is focused on reproducing the index-dependent broadening fingerprints of stacking disorder.

Because incomplete Ca-for-Na exchange could also broaden diffraction features, we assessed the cation composition by SEM-EDX. Figure 3 shows a representative spectrum together with an SEM image of the powder. In all measured spots, only Ca and Ir peaks (in addition to O) are observed, and no Na peak is detected above background at the expected Na K alpha energy (about 1.04 keV) within the instrumental detection limit. Semi-quantitative quantification (normalized to O = 3) over eight spots gives Ca = 0.848 (17) and Ir = 1.076 (9), corresponding to Ca/Ir = 0.788 (22). We note that the absolute Ca/Ir ratio obtained by EDX can be biased in a mixed light-element plus heavy-element oxide (Ca plus Ir) due to matrix effects and standard-based corrections, and therefore the Ca/Ir value should be regarded as semi-quantitative; a slight Ca deficiency cannot be strictly excluded.

Importantly, however, the absence of Na is a more robust conclusion than the absolute Ca/Ir ratio. For reference, a composition as Na-rich as $Ca_{0.8}Na_{0.4}IrO_3$ assuming $Ir^{4+}$ corresponds to about 4.40 wt% $Na_2O$, and Na at this several-wt% level would yield a clearly visible Na $K\text{-}\alpha$ feature near 1.04 keV, which is not observed. Therefore, the absence of detectable Na argues against a substantial residual Na-bearing fraction being the dominant origin of the anisotropic line broadening, and we therefore interpret the index-dependent

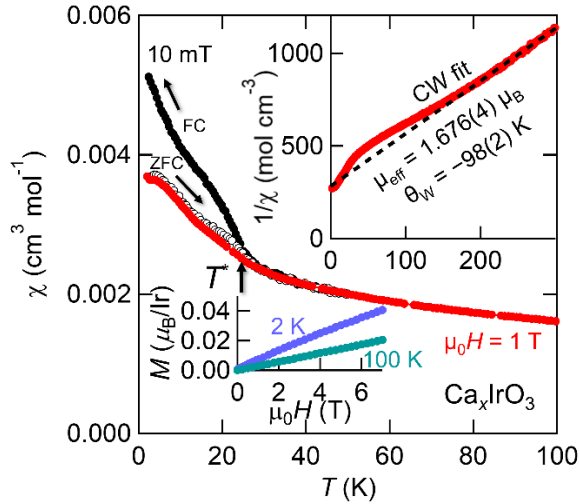
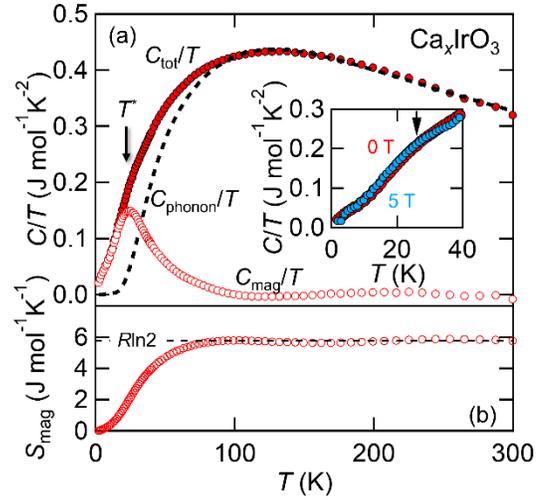

**Fig. 4** Temperature dependence of the magnetic susceptibility $\chi$ at 10 mT (filled: field-cooled; empty: zero-field-cooled) and 1 T (red). The upper inset shows the temperature dependence of the inverse susceptibility $1/\chi$; the lower inset shows isothermal magnetization $M$ at 2 K (purple) and 100 K (blue). The dashed line on the data is a Curie-Weiss fit.

**Fig. 5** (a) Temperature dependence of $C/T$ at 0 T. The black dashed line indicates the lattice contribution to $C/T$, estimated by fitting data above 100 K. The red circles are the magnetic specific heat obtained by subtracting the lattice contribution from the measured values. The inset shows the temperature dependence of $C/T$ at 0 T and 5 T. (b) Temperature dependence of the magnetic entropy calculated from the magnetic specific heat. The black dashed line marks the magnetic entropy for $J_{\text{eff}} = 1/2$, $S_{\text{mag}} = R \ln 2$.

broadening primarily in terms of layer-glide stacking disorder within the ilmenite framework.

Because many $ABO_3$ oxides prefer perovskite structures, it is useful to place stoichiometric $CaIrO_3$ in the empirical tolerance-factor landscape. The Goldschmidt perovskite factor [18],

$$t_p = \frac{r_A + r_X}{\sqrt{2}(r_B + r_X)} \quad (3)$$

correlates with perovskite stability. As summarized in Table II, perovskite-type $SrIrO_3$ and $BaIrO_3$ have $t_p \sim 1$ [46,47], whereas ilmenite-type $MgIrO_3$, $ZnIrO_3$, and $CdIrO_3$ fall in the range $0.74 \lesssim t_p \lesssim 0.82$ [24,26]. For $CaIrO_3$ we obtain $t_p = 0.96$, consistent with its well-known preference for perovskite-derived structures and with the ambient-pressure stability of the post-perovskite ($Cmcm$) phase over ilmenite [47]. Thus, from a simple size-metric viewpoint, $CaIrO_3$ is among the least expected compositions to support an ilmenite-type honeycomb framework at ambient pressure. We also quote, for reference, the ilmenite tolerance factor $t_i$ and the A–Ir electronegativity difference $\Delta e$ (Table III); these indicators suggest that $Ca_xIrO_3$ lies near the edge of ilmenite stability. In this context, the emergence of an ilmenite-type framework in $Ca_xIrO_3$ suggests that defect formation (stacking disorder) and A-site under-occupation can further facilitate trapping of the ilmenite topology at low temperature, even when the fully ordered stoichiometric $CaIrO_3$ ilmenite polymorph is not expected to be the ambient-pressure ground state. In other words, the present result highlights that access to an ilmenite-type polymorph in this unfavorable size regime is enabled, in practice, by low-temperature kinetic trapping together with pronounced stacking disorder.

### B. Physical Properties

Figure 4 shows the temperature dependence of the magnetic susceptibility, $\chi(T)$, for $Ca_xIrO_3$ measured under field-cooled (FC) and zero-field-cooled- (ZFC) conditions at $\mu_0 H = 1$ T and 10 mT. The upper inset plots the inverse susceptibility, $1/\chi$, and the lower inset displays isothermal $M(H)$ curves at 2 K and 100 K. At $\mu_0 H = 10$ mT, a gradual bifurcation between the FC and ZFC curves appears below $T^* \sim 25$ K, indicating a low-field, freezing-like magnetic anomaly with thermal hysteresis. In the high-temperature regime, we analyze $1/\chi$ within a Curie–Weiss framework; the values of effective moment $\mu_{\text{eff}} = 1.676(4)\ \mu_B$ and a Weiss temperature $\theta_W = -98(2)$ K are obtained over 200–300 K. Because the A-site content x in $Ca_xIrO_3$ is not uniquely fixed by SEM-EDX, there is a small systematic uncertainty in converting mass-normalized $\chi$ to a molar susceptibility. However, even if $x$ were as low as ~0.8, the formula mass would decrease by only about ~2.9%, and since $\mu_{eff}$ scales with the square root of the molar Curie constant, the resulting correction to $\mu_{eff}$ would be about ~1.4%; therefore, the following moment-based discussion is not qualitatively affected. The value of $\mu_{eff}$ is close to the 1.73 $\mu_B$ expected for an $Ir^{4+}$ $J_{eff} = 1/2$ doublet, implying that the local $IrO_6$ octahedral environment remains largely intact. Moreover, if $x$ were substantially below unity and charge compensation were dominated by oxidation of a significant fraction of $Ir^{4+}$ to $Ir^{5+}$ (expected to be $J_{eff} = 0$), $\mu_{eff}$ per Ir would be markedly reduced; the observed near-$J_{eff} = 1/2$ value therefore suggests that most Ir remains close to $Ir^{4+}$, consistent with $x$ being close to 1 and/or with charge compensation not proceeding primarily through $Ir^{5+}$ formation. Such a nearly ideal $J_{eff} = 1/2$ moment has also been reported for other honeycomb iridates, including $Na_2IrO_3$ and $Cu_2IrO_3$, even when substantial stacking disorder is present [49]. This similarity suggests that, in $Ca_xIrO_3$ as well, stacking faults primarily affect interlayer correlations while leaving the local $IrO_6$ crystal field and single-ion moment essentially unchanged. The negative Weiss temperature of $\theta_W \approx -98$ K points to dominant antiferromagnetic interactions on this $J_{eff} = 1/2$ network, and the absence of detectable hysteresis in the isothermal $M$ curves at 2 K and 100 K supports the lack of any sizable ferromagnetic component within our experimental resolution.

Figure 5 presents the temperature dependence of the specific heat divided by temperature, $C/T$. A weak hump appears near $T^* \sim 25$ K, coincident with the anomaly seen in $\chi(T)$. To quantify the entropy associated with this feature, the total specific heat was decomposed into lattice and magnetic parts, and the magnetic contribution $C_{mag}/T$ was obtained by subtracting the phonon term $C_{phonon}/T$ from $C_{tot}/T$.

The phonon contribution was modeled as the sum of a Debye term (three acoustic branches) and multiple Einstein terms (twelve optical branches; $3n - 3 = 12$ for $n = 5$ atoms per formula unit) [51]:

$$C_{lattice} = C_D + C_E = 9R\left(\frac{T}{\theta_D}\right)^3 \int_0^{\frac{\theta_D}{T}} \frac{x^4 \exp(x)}{[\exp(x)-1]^2} dx$$

$$+ R\sum_{i=1}^{4} n_i \frac{\left(\frac{\theta_{E_i}}{T}\right)^2 \exp\left(\frac{\theta_{E_i}}{T}\right)}{\left[\exp\left(\frac{\theta_{E_i}}{T}\right) - 1\right]^2}, \quad (6)$$

,where $R$ is the gas constant, and $\theta_D$ and $\theta_{Ei}$ denote Debye and Einstein temperatures, respectively. A fit performed above 100 K, where magnetic contributions are negligible, gives $\theta_D$

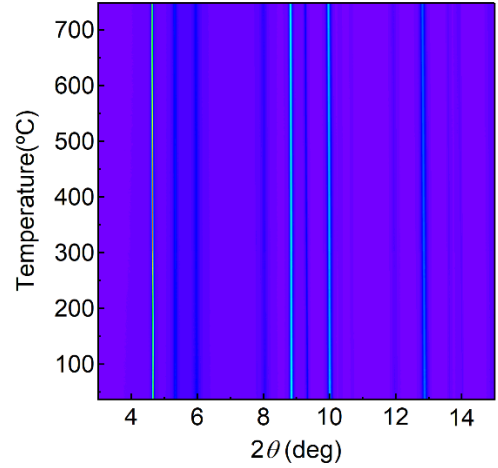

**Fig. 6** In-situ hot-stage synchrotron XRD heating profiles of ilmenite $Ca_xIrO_3$. The heating rate is 10 K/min. Diffraction intensity is shown as a temperature-dependent $2\theta$ map (color scale) collected during continuous heating at SPring-8 BL13XU [$\lambda$ = 0.413463Å]. The set of Bragg reflections assigned to ilmenite $CaIrO_3$ persists throughout the heating run up to about 750ºC, and no clear emergence of additional reflections attributable to decomposition products is observed within the sensitivity of the measurement.

= 452(8) K, $\theta_{E1}$ = 185(5) K, $\theta_{E2}$ = 305(5) K, $\theta_{E3}$ = 497(5) K, $\theta_{E4}$ = 1010(40) K, $n_1$ = 4, $n_2$ = 2, $n_3$ = 3, and $n_4$ = 3 (dashed line in Fig. 5).

The magnetic entropy $S_{mag}$ approaches $R \ln 2$ = 5.76 J $mol^{-1}$ $K^{-1}$ at high temperature, as expected for a $J_{eff} = 1/2$ doublet, supporting the validity of the phonon subtraction. The resulting $C_{mag}/T$ [see Fig. 4(a)] exhibits a broad but distinct maximum near $T^* \sim 25$ K. Upon applying magnetic fields up to 5 T, $C/T$ remains essentially unchanged in position within our resolution, while the anomaly becomes slightly sharper (narrower and of modestly larger amplitude). The associated entropy release is, within uncertainty, field-independent, indicating a redistribution over a narrower temperature window rather than a change in the number of magnetic degrees of freedom. This modest field sharpening contrasts with the canonical spin-glass response, where a magnetic field typically suppresses and broadens the anomaly, and instead suggests that the applied field moves $CaIrO_3$ closer to the ordered side of the phase boundary by quenching antiferromagnetic fluctuations and stabilizing the correlated state, in line with the general expectation and observations in frustrated magnets where magnetic fields sharpen thermodynamic signatures of order [52,53].

## IV. DISCUSSION

Beyond the known post-perovskite and perovskite forms, the present work expands the polymorphic family of $Ca_xIrO_3$ by establishing an ilmenite-type variant accessible at ambient pressure via low-temperature topochemical $Ca^{2+}/2Na^+$ exchange from $Na_2IrO_3$. Full-profile analysis of the powder SXRD diffraction data, together with stacking-fault refinements based on a first-order Markov stacking model, indicates pronounced stacking disorder that can be described as frequent switching faults between two symmetry-equivalent lateral stacking steps ($S_1$ and $S_2$) within the ilmenite-type framework. These $S_1/S_2$ switches introduce local AB-type registry (layer-glide fault motifs) into an otherwise rhombohedral ilmenite stacking sequence (ABC or its symmetry-equivalent variant), rather than implying a distinct long-range polytype. This behavior is consistent with a well ordered in plane arrangement and a fixed average interlayer spacing, while frequent in plane layer glides introduce correlated stacking faults between nominally equivalent layers. A key observation is that, despite this high fault density, a low field, freezing like bulk magnetic anomaly is observed near 25 K (ZFC/FC bifurcation and a broad $C/T$ maximum). We therefore avoid asserting unambiguous long range magnetic order based on the present dataset and instead interpret the results as dominant antiferromagnetic short-range correlations with defect induced freezing.

From a broader perspective, it is instructive to ask why an ilmenite-type polymorph of $Ca_xIrO_3$ can be accessed at all, given that many $ABO_3$ oxides prefer perovskite structures. Ilmenite stability can be discussed more specifically using an ilmenite tolerance factor $t_i$ together with a criterion based on the electronegativity difference between the $A$ ion and Ir, $\Delta e$ [46]:

$$t_i = \frac{1}{3}\left(\frac{(\sqrt{2}+1)r_O + r_B}{r_A + r_O} + \frac{\sqrt{2}r_O}{r_B + r_O}\right) > 0.80, \quad (4)$$
$$\Delta e > 1.465. \quad (5)$$

For $CaIrO_3$ $t_i \approx 0.96$ and $\Delta e = 1.20$ (Table III). Within these empirical criteria, $Ca_xIrO_3$ lies in a regime where the geometric size mismatch is still acceptable, but the electronegativity contrast between the $A$–O and Ir–O bonds is too small to provide a strong energetic driving force for the ordering characteristic of an ideal ilmenite lattice. In other words, if one only considers ionic size and electronegativity, a fully ordered ilmenite phase of $Ca_xIrO_3$ would not be expected to be the thermodynamic ground state at ambient pressure.

To complement this empirical view, we performed a 0 K convex-hull analysis based on DFT data [41]. The calculated energy distances to the hull are $\Delta H_{hull} = 0.0092$ eV/atom for post-perovskite and 0.014 eV/atom for ilmenite $CaIrO_3$. Thus, both polymorphs lie slightly above the hull and are formally metastable, and the nominal energy difference between them is only $\sim$0.005 eV/atom. This small offset is within typical DFT uncertainties and does not allow us to rank the two structures with high confidence. Nevertheless, the combined picture is consistent with experiment: the reduced electronegativity contrast means that the ilmenite lattice does not enjoy a strong thermodynamic advantage, so high-temperature solid-state synthesis tends to relax toward more stable competing phases or perovskite-derived networks, whereas low-temperature topochemical routes can bypass these equilibration pathways and kinetically trap the ilmenite-type $Ca_xIrO_3$ framework despite its marginal position above the convex hull.

To further assess the practical robustness of this metastable ilmenite-type $Ca_xIrO_3$, we performed time-resolved temperature-dependent synchrotron powder XRD during heating (see Fig. 6). In the measured low-angle window, the reflection set assigned to the ilmenite framework persists up to about $700\,^\circ$C without the emergence of additional peaks or abrupt changes that would signal decomposition or a reconstructive transformation to another polymorph. The peak positions shift smoothly with temperature, consistent with thermal expansion, while the characteristic line-shape features associated with stacking disorder remain qualitatively similar. This observation indicates that, although the ilmenite polymorph is only marginally metastable in a 0 K thermodynamic sense, it is protected by a sizable kinetic barrier under the present conditions. In practice, the phase is therefore sufficiently stable for bulk property measurements and for moderate thermal processing, and its accessibility is consistent with topochemical trapping of an inherited framework rather than equilibrium stabilization.

From a synthetic viewpoint, a smaller $\Delta e$ implies that the $A$–O and Ir–O bonds have similar ionicity/covalency, so placing $A$ and Ir in alternating layers (as in ilmenite) does not gain much additional binding energy over competing arrangements such as perovskite or post-perovskite. In such cases the ilmenite structure is disfavored under high-temperature, near-equilibrium ceramic conditions, where the system can more easily relax into phases with higher overall lattice stability. By contrast, low-temperature topochemical routes can exploit kinetic trapping: once the $IrO_6$ framework is inherited from $Na_2IrO_3$, $Ca^{2+}/2Na^+$ exchange can populate the A site without fully equilibrating the cation arrangement, allowing an ilmenite-type $Ca_xIrO_3$ framework to be preserved even though $\Delta e$ lies below the empirical stability threshold.

This interpretation is consistent with the behavior of other $A$IrO$_3$ ($A$ = Mg, Zn, Cd) ilmenites. As summarized in Table III, their $\Delta e$ values also fall below 1.465, and indeed these ilmenites have been obtained only by low-temperature or soft-chemistry routes and exhibit pronounced stacking faults. The recurring combination of reduced $\Delta e$, faulted ilmenite stacking, and the need for gentle synthetic conditions suggests that a smaller electronegativity difference not only lowers the thermodynamic stability of the ideal ilmenite structure but also enhances the role of defects (layer glides, mixed S1/S2 switching layer-glide stacking fault) as a practical means to stabilize metastable ilmenite-type iridates.

In honeycomb iridates with strong spin–orbit coupling, such as Ca$_x$IrO$_3$, the way the layers stack controls how IrO6 octahedra in neighboring sheets "see" each other and thus how magnetic interactions propagate in three dimensions [2-4]. In the ilmenite type CaIrO$_3$ studied here, our FAULTS analysis shows a random $S_1$/$S_2$ switching layer-glide stacking fault. The diffraction pattern indicates that the average spacing between layers is well defined, but the lateral registry from one layer to the next is locally disturbed. Within this structural context, the magnetic response can be viewed as the result of three related effects. First, layer glides randomize and weaken the interlayer magnetic couplings, so that correlations are strong within each honeycomb plane but less effective in locking different planes together [2-4]. Second, the local Ir–O–Ir geometry at glide faults slightly changes the balance among different exchange terms (Heisenberg, Kitaev type, and related anisotropic interactions), producing a distribution of interaction strengths that naturally broadens thermodynamic anomalies [2-4, 54]. Third, domain walls or twin-like boundaries associated with changes in lateral registry ($S_1$-dominated vs $S_2$-dominated stacking, and the intervening $S_1$/$S_2$ switching faults that generate AB-type registry) can host weakly coupled or uncompensated moments, which are prone to freezing under small dc fields [28-30,49]. Taken together, these features point to a freezing-like bulk magnetic anomaly arising from robust in plane correlations that do not achieve full three-dimensional coherence because the interlayer registry is disrupted by layer-glide faults, consistent with the negative Curie–Weiss $\theta_W$, the broad heat capacity maximum near $T^*$, and the low field ZFC/FC bifurcation without macroscopic hysteresis; accordingly, we do not assert long range magnetic order.

Comparison with related systems supports this view. In Ag$_3$LiIr$_2$O$_6$, long-range order is suppressed in highly faulted samples but reappears when stacking defects are reduced [28,29]. H$_3$LiIr$_2$O$_6$ contains large interlayer shifts and has been reported as a spin-liquid candidate, while the extent of defect-derived contributions to its continuum excitations and entropy remains under active discussion [30,39,40]. Ilmenite-type MgIrO$_3$ and ZnIrO$_3$ also host stacking faults; they do order magnetically, yet their heat-capacity anomalies are markedly broadened. Ca$_x$IrO$_3$ falls between these limits, suggesting that its ground state can be tuned by adjusting the fault density.

Thermodynamic considerations are consistent with this picture. As discussed above, 0 K convex-hull estimates place post-perovskite and stoichiometric ilmenite CaIrO$_3$ only a few meV per atom above the equilibrium hull [42]. Given this small energy scale and the typical uncertainties of DFT-based energetics, it is natural that mixed S1/S2 switching layer-glide stacking fault in the ilmenite framework, together with local strain relief at layer glides, can lower the effective free energy under low-temperature conditions. In this sense, stacking faults provide a route to defect-assisted stabilization of an ilmenite-type Ca$_x$IrO$_3$ polymorph without contradicting the high-temperature preference for perovskite-derived structures.

For the broader search for Kitaev physics, controlled introduction or suppression of stacking faults is a practical design variable. Because Ca$_x$IrO$_3$ sits in a critical window, it can be used to traverse from an ordered state, through a boundary regime, toward a disordered (spin-liquid-like) response by systematic control of stacking order. A central task moving forward is to disentangle responses that arise from intrinsic bond-anisotropic exchange from those caused by extrinsic structural defects. In several related honeycomb iridates, such as Ag$_3$LiIr$_2$O$_6$ and Cu$_2$IrO$_3$, improved synthetic control and reduced chemical disorder have been reported to favor the emergence of long-range magnetic order [28,55]. By analogy, it is plausible that Ca$_x$IrO$_3$ samples with lower stacking-fault density, obtained via further optimized topochemical protocols, would reside deeper on the ordered side of the phase diagram and could exhibit sharper thermodynamic signatures of magnetic ordering. More broadly, our results show that stacking disorder is not merely a source of sample "imperfection" but can act as a thermodynamic resource: layer glides help stabilize an ilmenite-type polymorph of Ca$_x$IrO$_3$ that lies close to, but above, the equilibrium convex hull and would be difficult to access by conventional high-temperature routes. This kind of defect-assisted stabilization provides a concrete strategy for expanding the structural phase space of spin–orbit-entangled oxides and for creating new honeycomb-lattice platforms that approach the ideal conditions for Kitaev physics and other exotic quantum ground states.

## V. SUMMARY

We synthesized metastable, ilmenite-type Ca$_x$IrO$_3$ by low-temperature topochemical ion exchange using Na$_2$IrO$_3$ as a precursor. Full-pattern SXRD analysis combined with stacking-fault refinements shows pronounced layer glide stacking fault that is well captured by a minimal model based on stochastic switching between two symmetry-equivalent lateral stacking steps ($S_1$ and $S_2$). This produces predominantly rhombohedral stacking sequences while intermittently introducing local AB-type registry and degrading lateral interlayer coherence. Magnetically, a freezing like anomaly appears near 25 K: the susceptibility exhibits a low field FC/ZFC bifurcation, and the specific heat shows a broad peak. These features point to dominant antiferromagnetic short-range correlations coexisting with defect induced glassy freezing, placing $Ca_xIrO_3$ near the boundary between the formation and suppression of long-range order. The ability to access this ilmenite-type framework despite the unfavorable crystal-chemistry of stoichiometric $CaIrO_3$ highlights the likely importance of defect formation (stacking disorder) together with *A*-site under-occupation in stabilizing the metastable polymorph.

This work was supported by JST PRESTO Grant Number JPMJPR23Q8 (Creation of Future Materials by Expanding Materials Exploration Space) and JSPS KAKENHI Grant Numbers. JP25K01496 (Scientific Research (B)), JP23H04616 and JP25H01649 (Transformative Research Areas (A) "Supra-ceramics"), JP24H01613 (Transformative Research Areas (A) "1000-Tesla Chemical Catastrophe"), JP25H01403 (Transformative Research Areas (B) "Multiply Programmed Layers"), JP22K14002 (Young Scientific Research), and JP24K06953 (Scientific Research (C)). Part of this work was carried out by joint research in the Institute for Solid State Physics, the University of Tokyo (Project Numbers 202311-MCBXG-0021, 202311-MCBXG-0025, 202406-MCBXG-0100, 202406-GNBXX-0095, 202406-MCBXG-0100, 202406-MCBXG-0101, 202411-MCBXG-0033, and 202411-MCBXG-0034). We thank S. Kobayashi and S. Kawaguchi in SPring-8 for assistance with the SXRD experiments. The synchrotron radiation experiments were performed at the Japan Synchrotron Radiation Research Institute, Japan (proposal No. 2025B1653).

## ACKNOWLEDGEMENT